\documentclass[a4paper,10pt]{article}
\usepackage[dvips]{color}
\usepackage{epsfig}
\usepackage{epsf}
\usepackage{amsmath}
\usepackage{graphicx}
\usepackage{float}
\usepackage{caption}
\usepackage{subcaption}

\begin{document}
\setcounter{page}{1}

\pagestyle{plain} \vspace{1cm}

\begin{center}
\Large{\bf Observational constraints on tachyonic chameleon dark energy model}\\
\small \vspace{1cm} {\bf A. Banijamali $^{a}$
\footnote{a.banijamali@nit.ac.ir}}, {\bf S. Bellucci $^{b}$
\footnote{Stefano.Bellucci@lnf.infn.it}}, {\bf B. Fazlpour
$^{c}$ \footnote{b.fazlpour@umz.ac.ir}} and {\bf M. Solbi $^{a}$}\\
\vspace{0.5cm} $^{a}$ {\it Department of Basic Sciences, Babol
Noshirvani University of Technology, Babol, Iran\\} \vspace{0.5cm}
$^{b}$ {\it INFN - Laboratori Nazionali di Frascati, 1-00044,
Frascati (Rome), Italy\\} \vspace{0.5cm} $^{c}$ {\it Department of
Physics, Babol Branch, Islamic Azad University, Babol, Iran\\}
\end{center}
\vspace{1.5cm}
\begin{abstract}
It has been recently shown that tachyonic chameleon model of dark
energy in which tachyon scalar field non-minimally coupled to the
matter admits stable scaling attractor solution that could give rise
to the late-time accelerated expansion of the universe and hence
alleviate the coincidence problem. In the present work, we use data
from Type Ia supernova (SN Ia) and Baryon Acoustic Oscillations to
place constraints on the model parameters. In our analysis we
consider in general exponential and non-exponential forms for the
non-minimal coupling function and tachyonic potential and show that
the scenario is compatible with observations.
\end{abstract}
\newpage
\section{Introduction}
Recent cosmological observations such as Type Ia supernova (SN
Ia)(Perlmutter et al., 1999; Riess et al., 1998), cosmic microwave
background (CMB) radiation (Ade et al., 2016; Ade et al., 2014;
Komatsu et al., 2011; Hinshaw et al., 2013), large scale structure
(Tegmark et al., 2004;  Seljak et al., 2005), Baryon Acoustic
Oscillations (BAO) (Eisenstein et al., 2005)  and weak lensing (Jain
\& Taylor, 2003) indicate that the expansion of the
current universe is accelerating.\\
Though the $\Lambda CDM$ cosmology in which $\Lambda$ is the
cosmological constant can successfully reproduce the late time
cosmic acceleration, it suffers from some serious issues such as the
fine tuning and cosmological coincidence problems (Ellis \& Madsen,
1995; Starobinsky, 1998). So, two alternative approaches have been
proposed to explain the current behaviour of the universe. The first
is to consider the modification of gravity on the large scale (see
(Nojiri \& Odintsov, 2007) for review and reference there in) and
the second is introducing dark energy sector in the content of the
universe (see (Copeland, Sami \& Tsujikawa, 2006) for review). It is
worthwhile to notice that in the second approach one can also
include a non-minimal coupling between dark energy and gravity to
construct scalar-tensor theories (see for example ( Nojiri \&
Odintsov, 2005)).\\
Besides scalar-tensor theories in which scalar field non-minimally
couples to the Ricci scalar or other geometric terms, recently
another form of a non-minimally coupled scalar field where the
scalar field is coupled to the matter has been widely studied in the
literature ( Das \& Banerjee, 2008;  Farajollahi et al., 2012;
Bisabr, 2012). This type of scalar field is known as a chameleon
field and thought it is heavy in the laboratory environment, on
cosmological scale where the matter density is small, the chameleon
is light enough to play the role of dark energy. \\
A chameleon scalar field has many remarkable cosmological features
as they pointed out by many authors. For instance, when it couples
to an electromagnetic field (Khoury \& Weltman, 2011)
\cite{Khoury:2011} in addition to the fluid then the fine tuning of
the initial condition on the chameleon may be resolved (Mota \&
Schelpe, 2012). Chameleon field can also successfully explain a
smooth transition from a deceleration to an acceleration epoch for
our universe (Banerjee \& Das, 2010). The scalar field that plays
the role of chameleon could be a Brans-Dicke scalar field with
interesting cosmological consequences such as explaining the current
accelerated expansion of the universe (Banerjee \& Das, 2010; Khoury
\& Weltman 2004; Das, Corasaniti \& Khoury, 2006). Koury in (Khoury,
2013) has summarized some important features of the chameleon field
theories. Moreover, phase-space analysis of chameleon scalar field
where a quintessence field acts as a chameleon has been investigate
in (Roy \& Banerjee, 2015). We have also studied the dynamics of
chameleon model where tachyon field plays the role of chameleon
(Banijamali \& Solbi, 2017). We have utilized the dynamical system
tools to obtain the critical points of such
theory and showed its interesting cosmological behaviour.\\
In the present work we apply combined datasets of Type Ia Supernova
(SN Ia) and Baryon Acoustic Oscillations to test the tachyonic
chameleon model and constrain its parameters. We use $\chi^2$
minimization technique to find the best fit values of the model
parameters and to plot the likelihood contours for them.\\
The paper is organized as follows: In the next section we have
presented the tachyonic chameleon model and derived the basic
equations along with the definitions of different cosmological
parameters relevant for our study. In section 3 we have briefly
discussed our methodology to use data from SN Ia and BAO
observations. In section 4 we have used the observational data to
plot likelihood contours of the model for different categories of
scalar potential and coupling function. Section 5 is devoted to our
conclusion.\\
\section{Basic equations}
Tachyonic chameleon model of dark energy in the framework of general
relativity can be described by the following action

\begin{equation}\label{action}
S=\int d^{4}x \sqrt{-g} \Big[\frac{R}{16\pi G}
-V(\phi)\sqrt{1-\partial_{\mu} \phi \partial^{\nu}
\phi}+f(\phi)\mathcal{L}_{m}\Big],
\end{equation}\\
where $g$ is the determinant of the metric tensor $g_{\mu\nu}$, $R$
is the Ricci scalar and $G$ is the bare gravitational constant.
$f(\phi)$ is the coupling function between scalar field and the
matter. $\mathcal{L}_{m}$ is the matter Lagrangian and $V(\phi)$ is
also the tachyon potential. \\
Variation of action (\ref{action}) with respect to the metric
$g_{\mu\nu}$ leads to the gravitational field equations. These
equations in FRW background with the metric

\begin{equation}\label{metric}
ds^{2}=dt^{2}-a^{2}(dr^{2}+r^{2}d\theta^{2}+r^{2}sin^{2}\theta
d\phi^{2}),
\end{equation}
take the following forms:

\begin{equation}\label{Friedman1}
3H^2=\rho_{m} f+\frac{V(\phi)}{\sqrt{1-\dot{\phi}^2}},
\end{equation}

\begin{equation}\label{Friedman2}
2\dot{H}+3H^2=-\gamma\rho_{m} f +V(\phi)\sqrt{1-\dot{\phi}^2},
\end{equation}
where $\rho_{m}$ and $p_m$ are the matter energy density and
pressure respectively. In fact equations (\ref{Friedman1}) and
(\ref{Friedman2}) are the Friedmann equations for our model.
Note also that  $p_{m}=\gamma \rho_{m}$ is assumed in deriving these equations.\\
On the other hand, variation of (\ref{action}) with respect to
thescalar field in FRW background yields to
\begin{equation}\label{fieldeqution}
\ddot{\phi}+(1-\dot{\phi}^2)(3H\dot{\phi}+\frac{V'}{V})=(1-3\gamma)
(1-\dot{\phi}^{2})^{\frac{3}{2}} \frac{\rho_{m}f'}{V},\end{equation}\\
where the tachyon field is assumed to be homogeneous and a prime
stands for derivative with respect to the $\Phi$.\\
In addition, the continuity equation reads,

\begin{equation}\label{continuity}
(\rho_{m}f)\dot{
}+3H(1+\gamma)\rho_{m}f=-(1-3\gamma)\rho_{m}\dot{f}.
\end{equation}\\
Now, we mention that equation (\ref{continuity}) has a solution as
follows:
\begin{equation}\label{solution}
\rho_{m}=\rho_{0} a^{-3(1+\gamma)}f^{-(2-3\gamma)}
\end{equation}\\
where $\rho_{0}$ is an integration constant. This shows the
evolution of the matter density strongly depends on the coupling
function $f$. When $f=1$, this solution reduces to the standard
evolution law for the matter energy density.\\
Furthermore, one can define the effective equation of state as
\begin{equation}\label{EOS}
\omega_{eff}\equiv\frac{p_{eff}}{\rho_{eff}},
\end{equation}\\
where $p_{eff}$ and $\rho_{eff}$ can be obtained from
(\ref{Friedman1}) and (\ref{Friedman2}) as follows:
\begin{equation}\label{peff}
p_{eff}=\gamma\rho_{m}f-V(\phi)\sqrt{1-\dot{\phi}^2},
\end{equation}
\begin{equation}\label{rhoeff}
\rho_{eff}=\rho_{m}f+\frac{V(\phi)}{\sqrt{1-\dot{\phi}^2}}.
\end{equation}
Before closing this section notice that, tachyonic chameleon dark
energy model exhibits some interesting cosmological implication from
dynamical system point of view. Depending on the form of coupling
function $f(\phi)$ and potential $V(\phi)$ such a model provides a
solution to coincidence problem and explains the current phase of
accelerated expansion of the universe ( Banijamali \& Solbi 2017).
Therefore, investigating this scenario using observational cosmology
and constraining the model parameters according to the latest data
is not only interesting but also necessary. This study will be done
in the next sections.\\
\section{Observational constraints on the model parameters}
In this section, we will fit the model parameters with recent
observational data from Type Ia Supernova (SN Ia) and Baryon
Acoustic Oscillations (BAO) observations. Although a well-known
analysis method is used in the literature, we briefly explain the
method for
the elaboration of the observational data.\\
The total $\chi^2$ for combined data analysis is given by:
\begin{equation}
\chi^{2}=\chi_{\mathrm{SN}}^{2}+\chi_{\mathrm{BAO}}^{2},
\label{eq:A.21}
\end{equation}
where each $\chi^2$ will be evaluated individually. We mention that
in our fitting method we use the simple $\chi^{2}$
method~(Perivolaropoulos 2013) , rather than the Markov-chain Monte
Carlo (MCMC) procedure such as CosmoMC~(Lewis \& Bridle 2002).\\
Now we are going to explain the way by which one can calculate each
$\chi^2$ (Yang et al 2010; Li et al 2010) .
\subsection{Type Ia Supernova (SN Ia)}
First, we have used the latest observational dataset of SN Ia which
give the information on the luminosity distance $D_L$ as a function
of the red shift $z$.\\
The Hubble-free luminosity distance for the flat universe is defined
as
\begin{equation}
D_{L}(z)=\left(1+z\right)\int_{0}^{z}\frac{dz'}{E(z')}\,,
\label{eq:A.2}\nonumber
\end{equation}
where $E(z) \equiv H(z)/H_{0}$,
with
{\small{
\begin{equation}
\frac{H(z)}{H_{0}}=
\sqrt{\Omega_{\mathrm{m}}^{(0)}\left(1+z\right)^{3}
+\Omega_{\mathrm{r}}^{(0)}\left(1+z\right)^{4}
+\Omega_{\mathrm{DE}}^{(0)}\left(1+z\right)^{3\left(1+w_{
\mathrm { DE } }
\right)}
}\,.
\label{eq:A.3}\nonumber
\end{equation}}}
Here, $\Omega_\mathrm{r}$ is the radiation density parameter and
$\Omega_{\mathrm{r}}^{(0)}=\Omega_{\gamma}^{(0)}
\left(1+0.2271N_{\mathrm{eff}}\right)$, where
$\Omega_{\gamma}^{(0)}$ is the present fractional photon energy
density and $N_{\mathrm{eff}}=3.04$ is the effective number of
neutrino species~(Komatsu et al 2011).\\
The theoretical distance modulus $\mu_{\mathrm{th}}$ is defined by
\begin{equation}
\mu_{\mathrm{th}}(z_{i})\equiv5\log_{10}D_{L}(z_{i})+\mu_{0}\,,
\label{eq:A.1}\nonumber
\end{equation}
where $\mu_{0}\equiv42.38-5\log_{10}h$, with $h \equiv
H_{0}/100/[\mathrm{km} \, \mathrm{sec}^{-1} \, \mathrm{Mpc}^{-1}]$
(Komatsu et al 2011).\\
For SN Ia dataset, $\chi^2$ function is given by
\begin{equation}
\chi_{\mathrm{SN}}^{2}=\sum_{i}\frac{\left[\mu_{\mathrm{obs}}(z_{i})-
\mu_{\mathrm{th}}(z_{i})\right]^{2}}{\sigma_{i}^{2}}\,,
\label{eq:A.4}
\end{equation}
where $\mu_{\mathrm{obs}}$ is the observed value of the distance modulus.
$\chi_{\mathrm{SN}}^{2}$ in the above equation can by expanded as (
Perivolaropoulos 2005)
\begin{equation}
\chi_\mathrm{SN}^{2}=A-2\mu_{0}B+\mu_{0}^{2}C\,,
\label{eq:A.5}\nonumber
\end{equation}
where
\begin{eqnarray}
&&A
=
\sum_{i}\frac{\left[\mu_{\mathrm{obs}}(z_{i})-\mu_{\mathrm{th}}(z_{i};\mu_{
0}=0)\right]^{2}}{\sigma_{i}^{2}}\,,
\quad\nonumber\\
&&B
=
\sum_{i}\frac{\mu_{\mathrm{obs}}(z_{i})-\mu_{\mathrm{th}}(z_{i};\mu_{0}=0)}
{\sigma_{i}^{2}}\,,
\quad\nonumber\\
&&C=\sum_{i}\frac{1}{\sigma_{i}^{2}}\,.
\label{eq:A.6}\nonumber
\end{eqnarray}
Note that $\mu_{obs}$ and $\mu_{th}$ represent the observed and
theoretical distance modulus respectively. $\sigma_i$ is also the
uncertainly in the distance modulus.\\
Finally, minimizing $\chi_{\mathrm{SN}}^{2}$  with respect to
$\mu_{0}$ leads to
\begin{equation}
\tilde{\chi}_{\mathrm{SN}}^{2}=A-\frac{B^{2}}{C}\,.
\label{eq:A.7}
\end{equation}
We use (\ref{eq:A.7}) for minimization of $\chi^{2}$ for 580 recent
data points of SN Ia (Suzuki et al. 2011).
\\\\
\subsection{ Baryon Acoustic Oscillations (BAO)}
We have also used BAO dataset to constrain our model parameters. The
distance ratio $d_{z}\equiv r_{s}(z_{\mathrm{d}})/D_{V}(z)$ is
measured by BAO observations. Here $D_{V}$ is the volume-averaged
distance, $r_{s}$ is the comoving sound horizon and $z_{\mathrm{d}}$
is the redshift at the drag epoch~(Percival et al. 2009).\\
Distance $D_{V}(z)$ is defined as~(Eisenstein 2005)
\begin{equation}
D_{V}(z)\equiv\left[\left(1+z\right)^{2}
D_{A}^{2}(z)\frac{z}{H(z)}\right]^{1/3}\,,
\label{eq:A.8}\nonumber
\end{equation}
where $D_{A}(z)$ is the proper angular diameter distance for the flat
universe.\\
In our analysis we use the 6dF, the SDSS and WiggleZ BAO data points
which are represented in Table \ref{tab1}. The WiggleZ collaboration
(Blake et al. 2011) has measured the baryon acoustic scale at three
different redshifts, while SDSS and 6DFGS surveys provide data at
lower redshift (Percival 2010).\\
\begin{table}[H]
\begin{center}
\begin{tabular}{| c | c | cc | ccc |}
\multicolumn{1}{c}{} & \multicolumn{1}{c}{6dF} & \multicolumn{2}{c}{SDSS}
 & \multicolumn{3}{c}{WiggleZ}  \\\hline
$z$ & 0.106 & 0.2 & 0.35 & 0.44 & 0.6 & 0.73  \\\hline
$d_z$ & 0.336 & 0.1905 & 0.1097 & 0.0916 & 0.0726 & 0.0592  \\\hline
$\Delta d_z$ & 0.015 & 0.0061 & 0.0036 & 0.0071 & 0.0034 & 0.0032 \\\hline
 \end{tabular}
 \caption{The BAO data used in our analysis.\label{tab1}}
\end{center}
\end{table}
\hspace*{-1cm}In addition, $C_{\mathrm{BAO}}^{-1}$ was obtained from
the covariance data (Blake et al. 2011) in terms of $d_z$ as
follows:
%
\begin{equation}
C_{\mathrm{BAO}}^{-1}=\left(
\begin{array}{cccccc}
4444&0&0&0&0&0\\
0&30318&-17312&0&0&0\\
0&-17312&87046&0&0&0\\
0&0&0&23857&-22747&10586\\
0&0&0&-22747&128729&-59907\\
0&0&0&10586&-59907&125536
\end{array}\right)\,.
\label{eq:A.13}\nonumber
\end{equation}
%
At last, $\chi^{2}$ for the BAO data is expressed as
%
\begin{equation}
\chi_{\mathrm{BAO}}^{2}=
\left(x_{i,\mathrm{BAO}}^{\mathrm{th}}-x_{i,\mathrm{BAO}}^{\mathrm{obs}}
\right)
\left(C_{\mathrm{BAO}}^{-1}\right)_{ij}
\left(x_{j,\mathrm{BAO}}^{\mathrm{th}}-x_{j,\mathrm{BAO}}^{\mathrm{obs}}
\right),
\label{eq:A.14}\nonumber
\end{equation}
%
where the indices $i,j$ are in growing order in $z$, as in Table
\ref{tab1}. In the next section we use the above two sets of data to
examine our model.\\
\section{Observational constraints on the model parameters}
Two important functions in the model (\ref{action}) are the
chameleon field potential $V(\phi)$ and coupling function $f(\phi)$.
We perform our analysis based on this fact that whether these
functions are exponential (power-law) or not. Thus we have four
categories given by the following subsections:
\subsection{Exponential $f(\phi)$ and power-law $V(\phi)$}
\begin{figure}[H]
    \centering
    \begin{subfigure}[H]{0.3\textwidth}
        \centering
        \includegraphics[width=50mm]{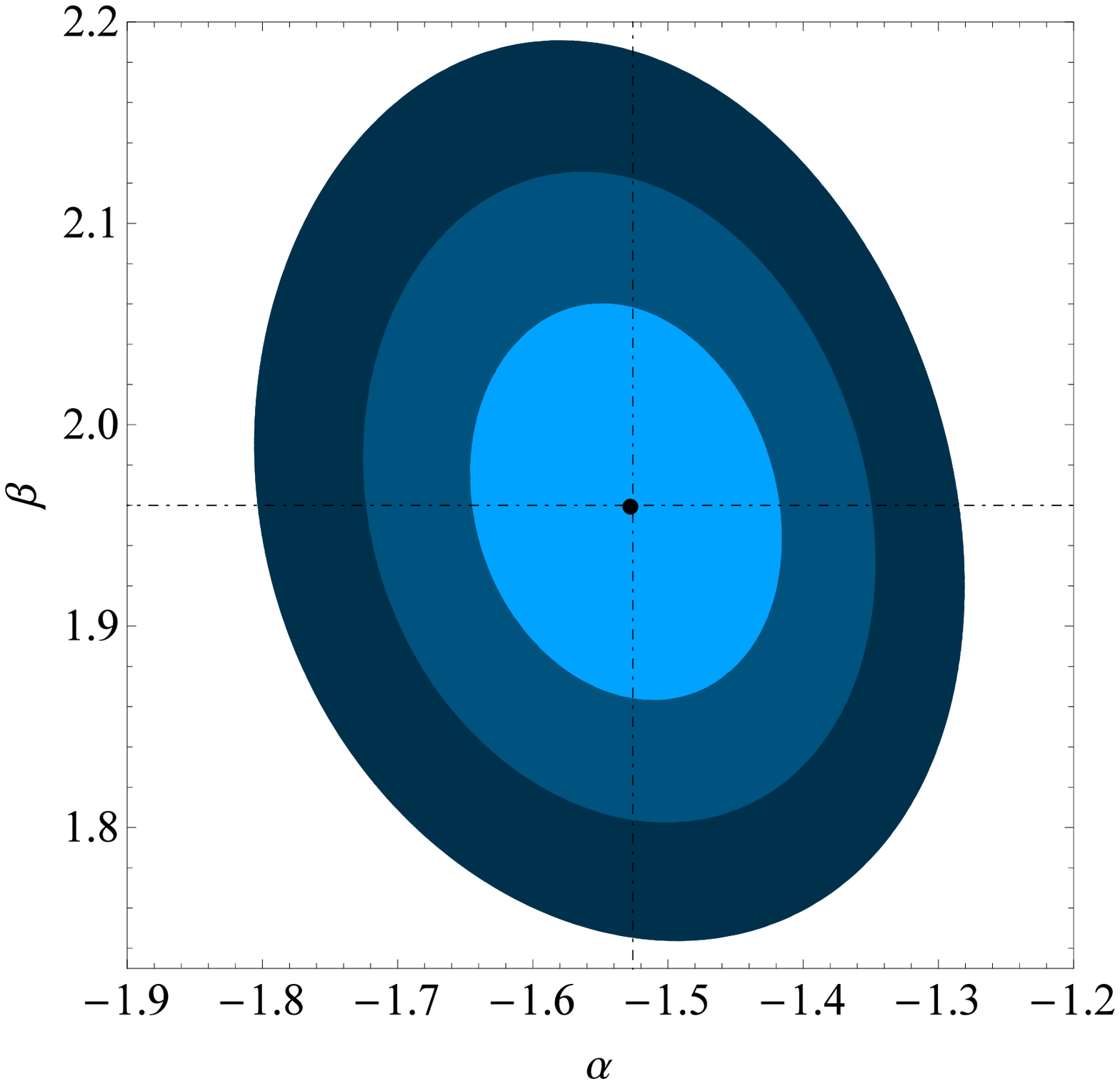}
        \caption{}
        \label{alphabeta}
    \end{subfigure}
    \begin{subfigure}[H]{0.3\textwidth}
        \centering
        \includegraphics[width=50mm]{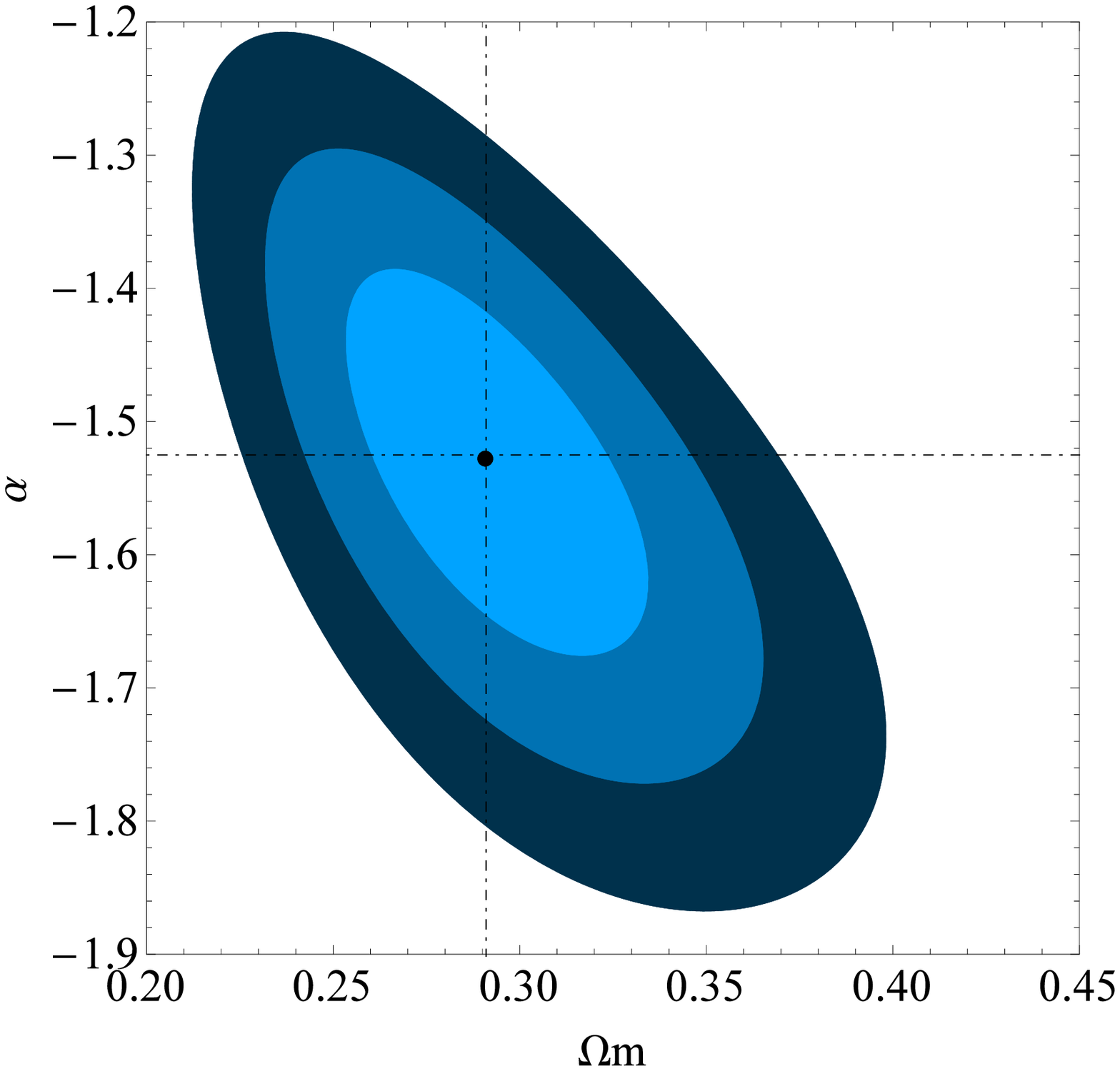}
        \caption{}
        \label{betaomega}
    \end{subfigure}
       \begin{subfigure}[H]{0.3\textwidth}
        \centering
        \includegraphics[width=50mm]{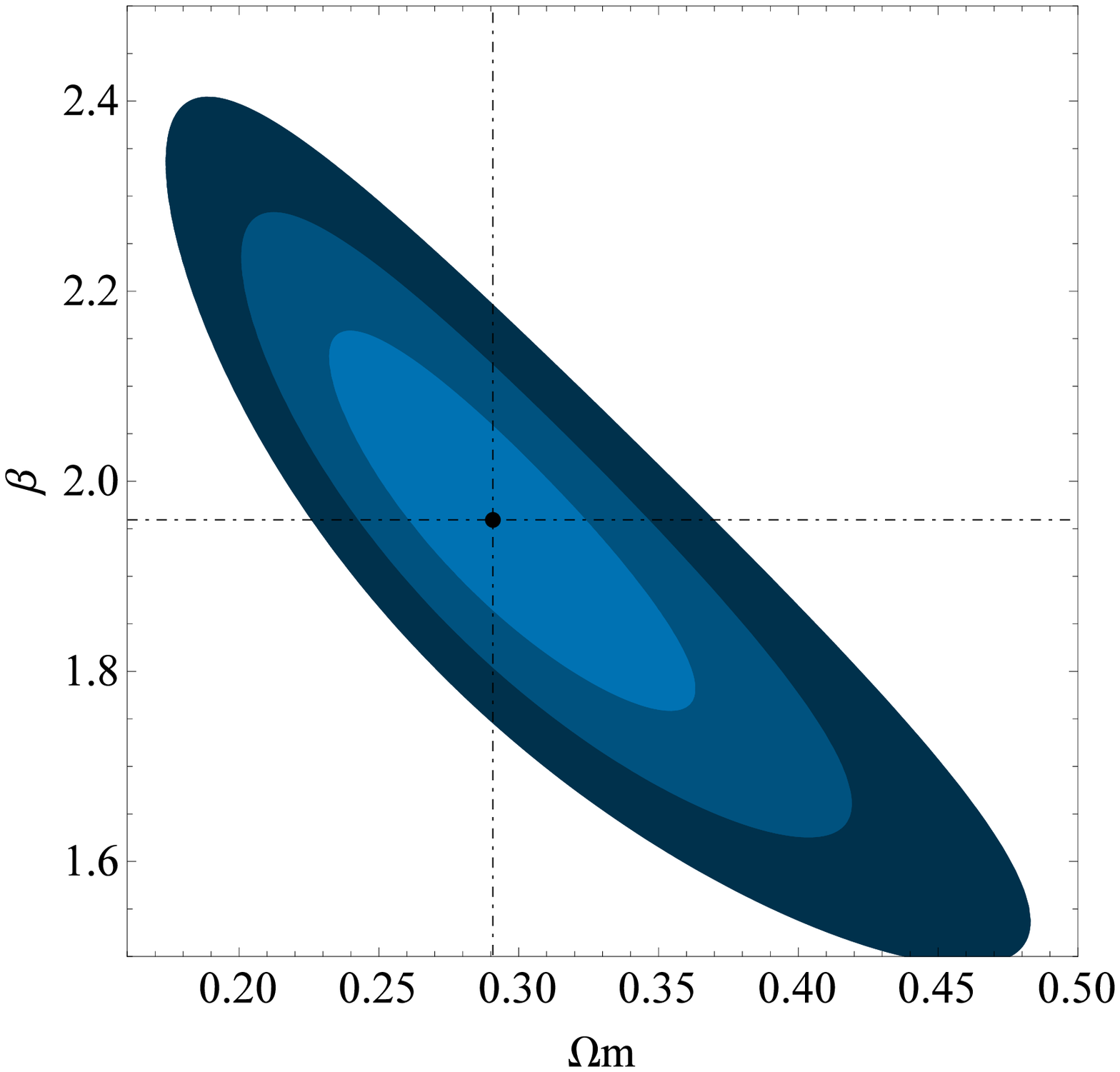}
        \caption{}
        \label{}
    \end{subfigure}
        \caption{Plots of $1\sigma$ (light blue), $2\sigma$ (blue) and $3\sigma$
        (dark blue) confidence contours on $\alpha-\beta$ (a), $\alpha-\Omega_{m_{0}}$ (b)
        and $\beta-\Omega_{m_{0}}$ (c) parameter spaces for SN Ia+BAO datasets in tachyonic
        chameleon dark energy scenario with $f(\phi)=f_0 e^{\alpha\phi}$ and
        $V(\phi)=V_{0} \phi^\beta$. The black dot in each plot shows the best fit point.}
    \label{fexp}
\end{figure}

In the first case we consider an exponential coupling function and a
power-law potential as follows'
\begin{align}
f(\phi) &= f_{0} e^{\alpha\phi},\nonumber\\
V(\phi) &= V_{0} \phi^\beta.
\end{align}
We are interested in constraining model parameters $\alpha$ and
$\beta$ together with the present values of the density parameters
using  the $\chi^2$-method for recent observational data. Thus we
plot the likelihood contours for these physically important
parameters and obtain the best fit values. We produce the likelihood
contours for $1\sigma$ , $2\sigma$ and $3\sigma$ confidence
levels.\\
\begin{table}[H]
\begin{center}
\begin{tabular}{|c||c||c||c||c|}
\hline Data&$\chi_{min}^2$&$\Omega_{m_{0}}$&$\alpha$&$\beta$\\
\hline
SN Ia +&584.08&0.29&-1.53&1.96\\
BAO&&&&\\
\hline
\end{tabular}
\caption{The value of $\chi_{min}^2$ and the best fit values of
model parameters $\Omega_{m_{0}}$, $\alpha$ and $\beta$ for the
first case.} \label{tab2}
\end{center}
\end{table}
In Figure \ref{fexp} the $1\sigma$, $2\sigma$ and $3\sigma$
confidence level contours for $\alpha-\beta$ (a),
$\alpha-\Omega_{m_{0}}$ (b) and $\beta-\Omega_{m_{0}}$ (c) are
plotted for SN Ia +BAO datasets. For simplicity reasons we set
$f_{0}=V_{0}=1$. As one can see in this figure this model is in
agreement with observations. The best-fit values of $\alpha$,
$\beta$ and $\Omega_{m_{0}}$ in addition to $\chi^{2}_{min}$ (the
minimum value of chi-square) are presented in Table \ref{tab2}. Note
that the best fit value of $\Omega_{m_{0}}=0.29$
is consistent with observations.\\
\subsection{Power-law $f(\phi)$ and exponential $V(\phi)$}
\begin{figure}[H]
    \centering
    \begin{subfigure}[H]{0.3\textwidth}
        \centering
        \includegraphics[width=50mm]{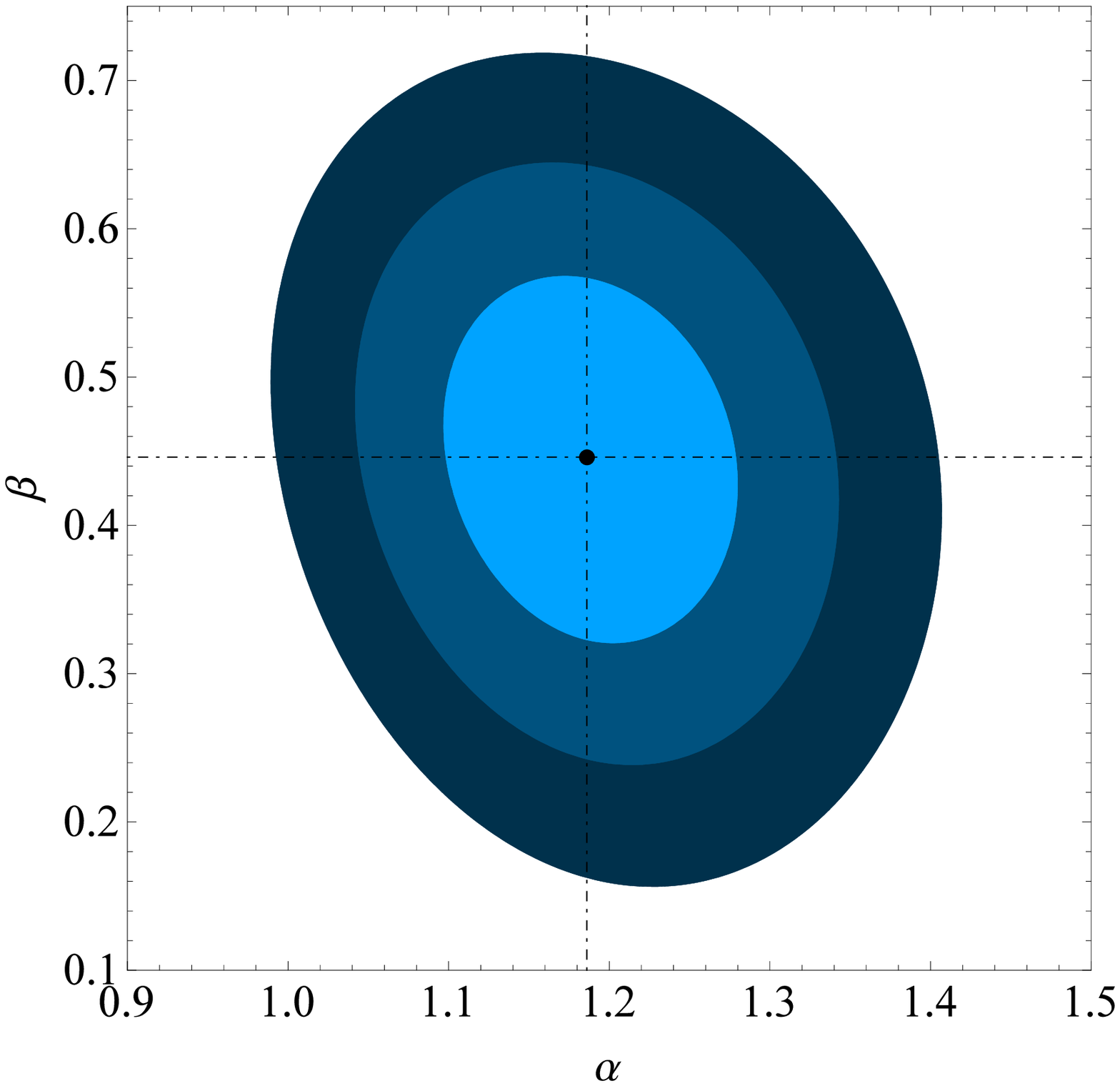}
        \caption{}
        \label{per:x-N}
    \end{subfigure}
    \begin{subfigure}[H]{0.3\textwidth}
        \centering
        \includegraphics[width=50mm]{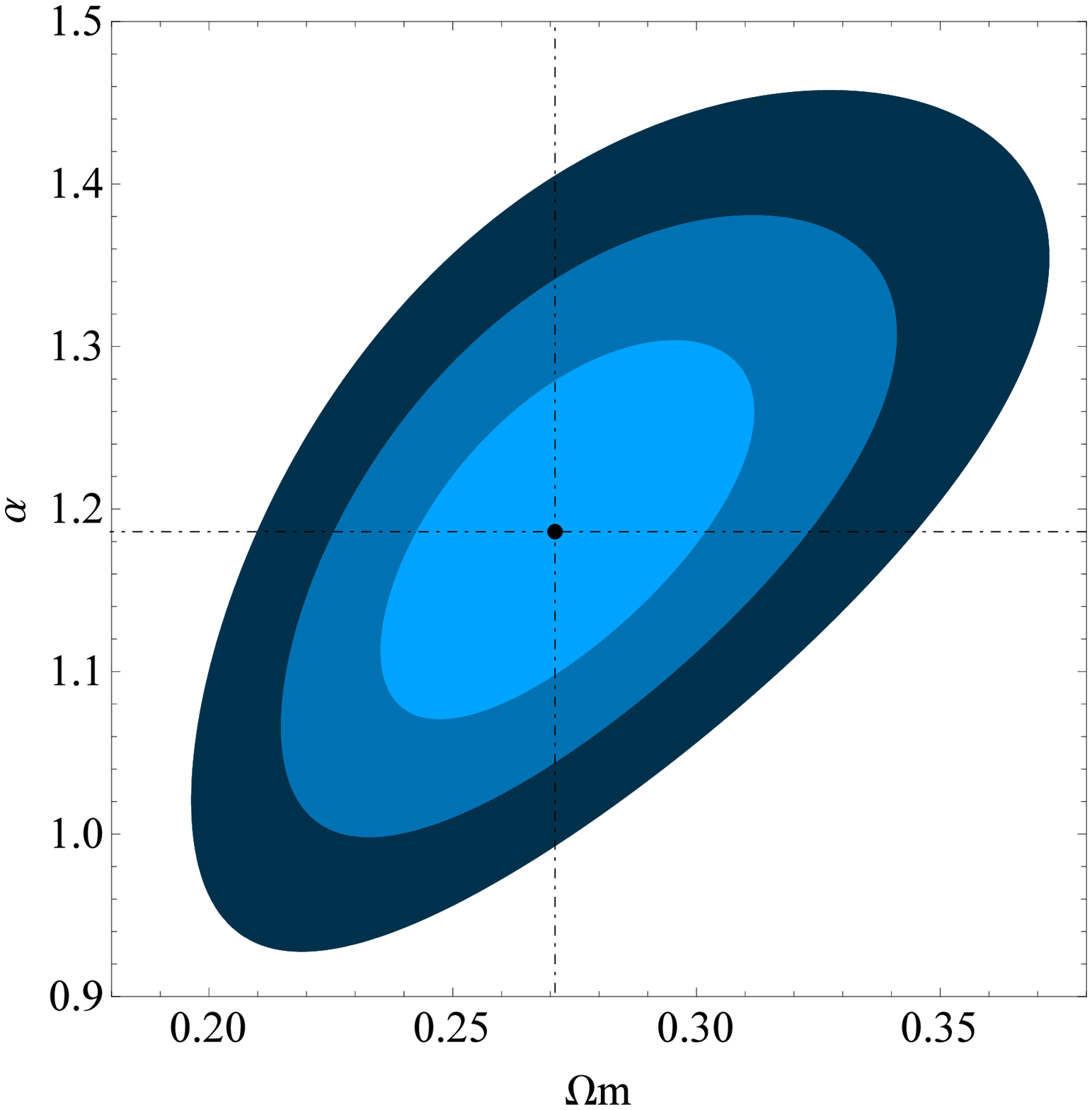}
        \caption{}
        \label{per:z-N}
    \end{subfigure}
       \begin{subfigure}[H]{0.3\textwidth}
        \centering
        \includegraphics[width=50mm]{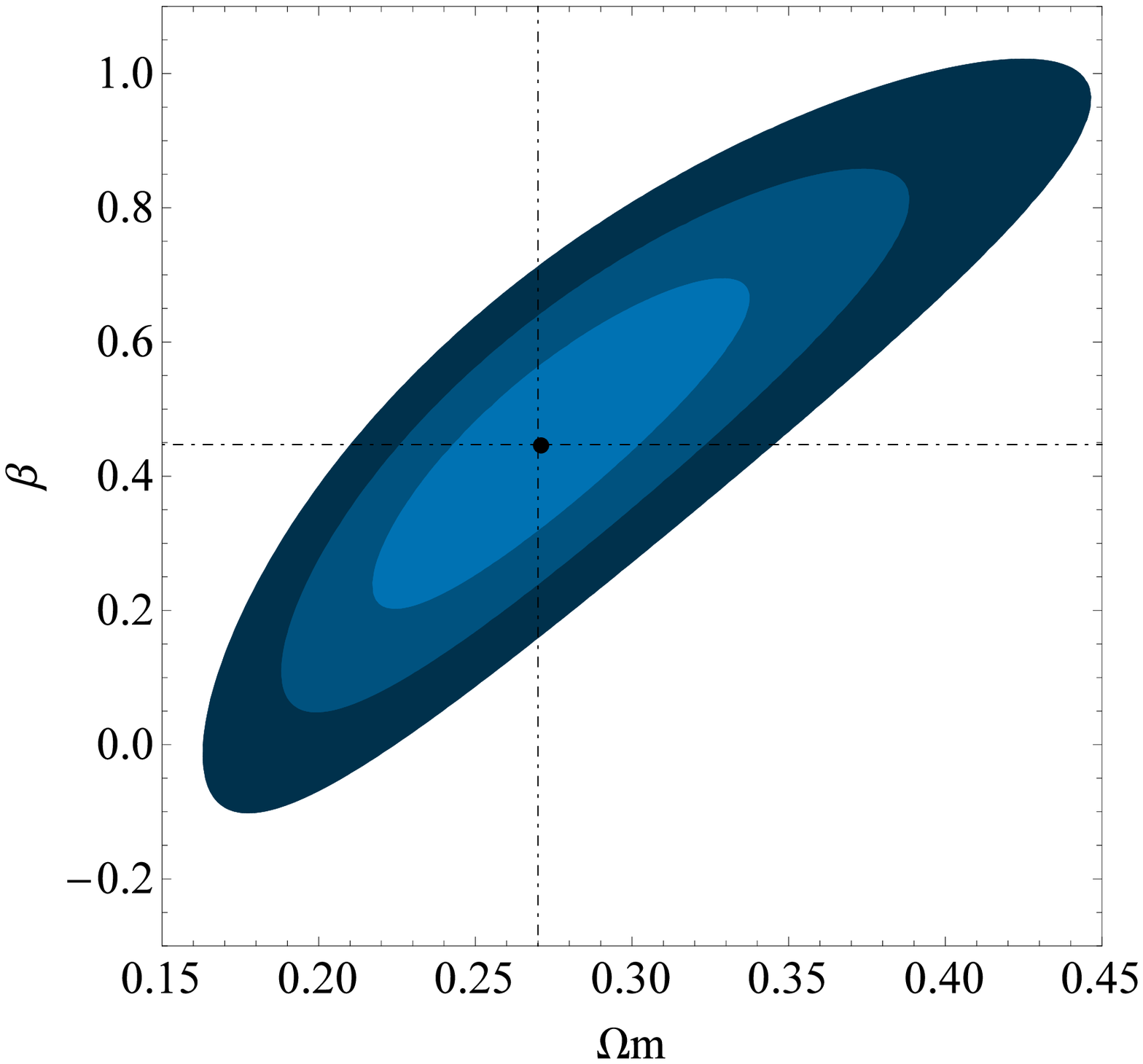}
        \caption{}
        \label{per:u-N}
    \end{subfigure}
        \caption{Plots of $1\sigma$ (light blue), $2\sigma$ (blue) and
        $3\sigma$ (dark blue) confidence contours on $\alpha-\beta$ (a),
        $\alpha-\Omega_{m_{0}}$ (b) and $\beta-\Omega_{m_{0}}$ (c)
        parameter spaces for SN Ia+BAO dataset in tachyonic chameleon dark
        energy scenario with $f(\phi)=f_0 \phi^{\alpha}$ and $V(\phi)=V_{0}
        e^{\beta\phi}$. The black dot in each plot shows the best fit point. }
    \label{vexp}
\end{figure}

For the second case we are going to constrain the model parameters for an exponential potential and a power-law $f(\phi)$ given by,
\begin{align}
f(\phi) &= f_{0} \phi^\alpha,\nonumber\\
V(\phi) &= V_{0} e^{\beta\phi} .
\end{align}
The best fit values of the model parameters $\alpha$ , $\beta$ and
$\Omega_{m_{0}}$ have been shown in Table \ref{tab3}. The
corresponding contours for $1\sigma$, $2\sigma$ and $3\sigma$
confidence level on $\alpha-\beta (a)$, $\alpha-\Omega_{m_{0}} (b)$
and $\beta-\Omega_{m_{0}} (c)$ planes are also plotted in Figure
\ref{vexp}. For simplicity reasons we set $f_{0}=V_{0}=1$. It is
interesting that these results are in good agreement with the
observational data. It deserves mention here that the value of
$\Omega_{m_{0}}$ obtained in the present paper is very close to the
expected value.
\begin{table}[H]
\begin{center}
\begin{tabular}{|c||c||c||c||c|}
\hline
Data&$\chi_{min}^2$&$\Omega_{m_{0}}$&$\alpha$&$\beta$\\
\hline
SN Ia +&584.08&0.270&1.18&0.44\\
BAO&&&&\\
\hline
\end{tabular}
\caption{The value of $\chi_{min}^2$ and the best fit values of
model parameters $\Omega_{m_{0}}$, $\alpha$ and $\beta$ for the
second case.} \label{tab3}
\end{center}
\end{table}

\subsection{Power-law $f(\phi)$ and $V(\phi)$}
\begin{figure}[H]
    \centering
    \begin{subfigure}[H]{0.3\textwidth}
        \centering
        \includegraphics[width=50mm]{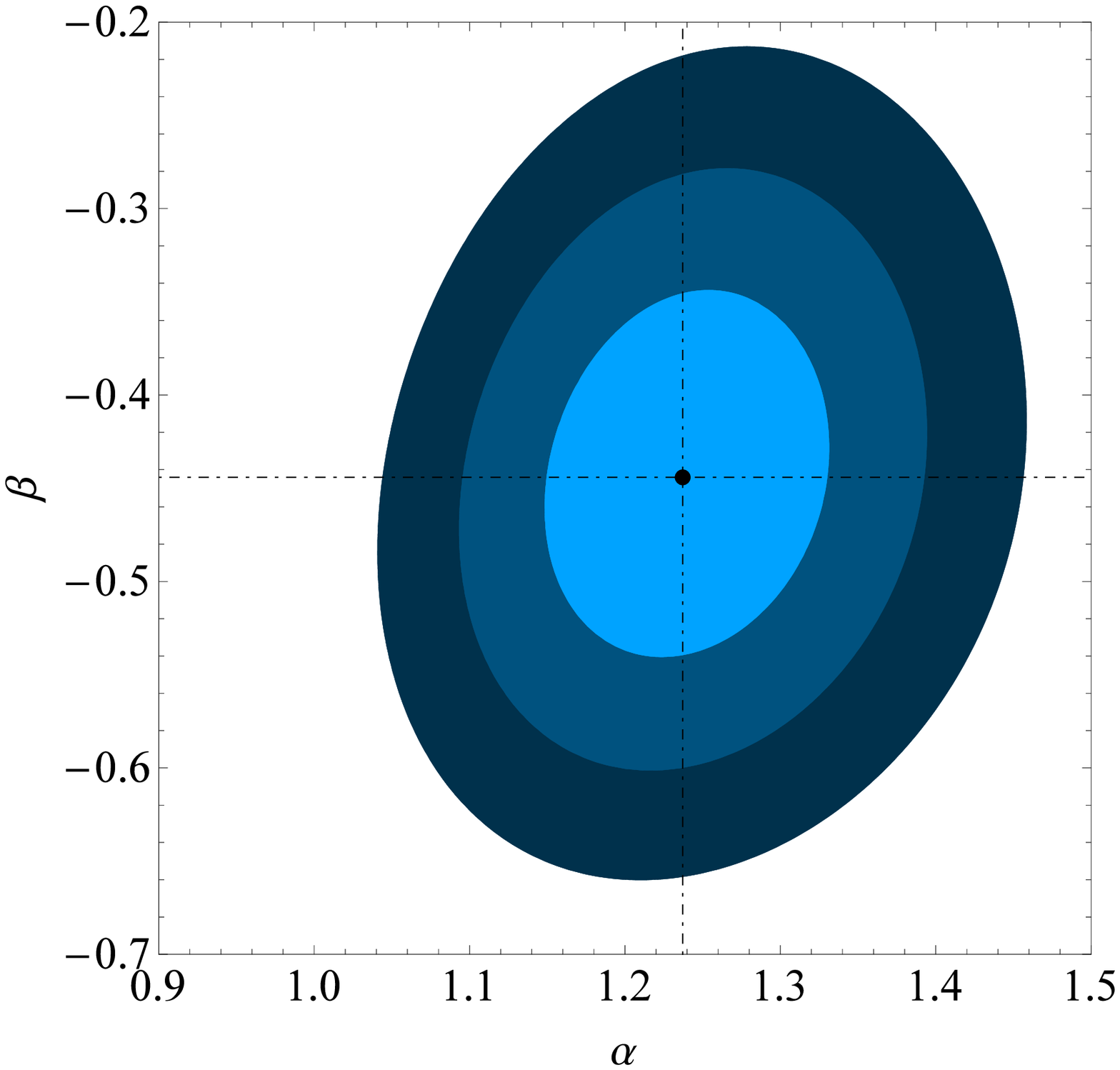}
        \caption{}
        \label{per:x-N}
    \end{subfigure}
    \begin{subfigure}[H]{0.3\textwidth}
        \centering
        \includegraphics[width=50mm]{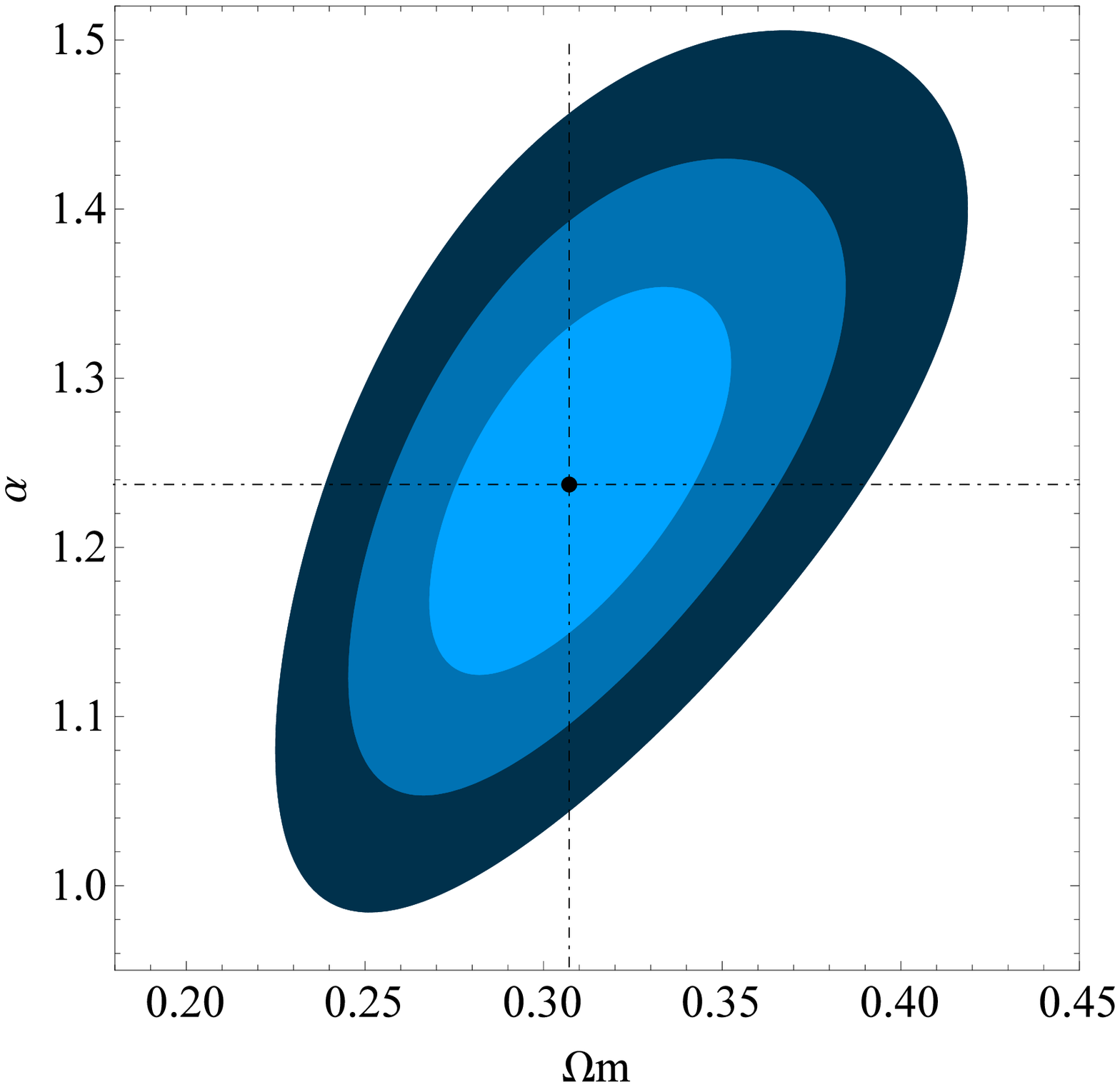}
        \caption{}
        \label{per:z-N}
    \end{subfigure}
       \begin{subfigure}[H]{0.3\textwidth}
        \centering
        \includegraphics[width=50mm]{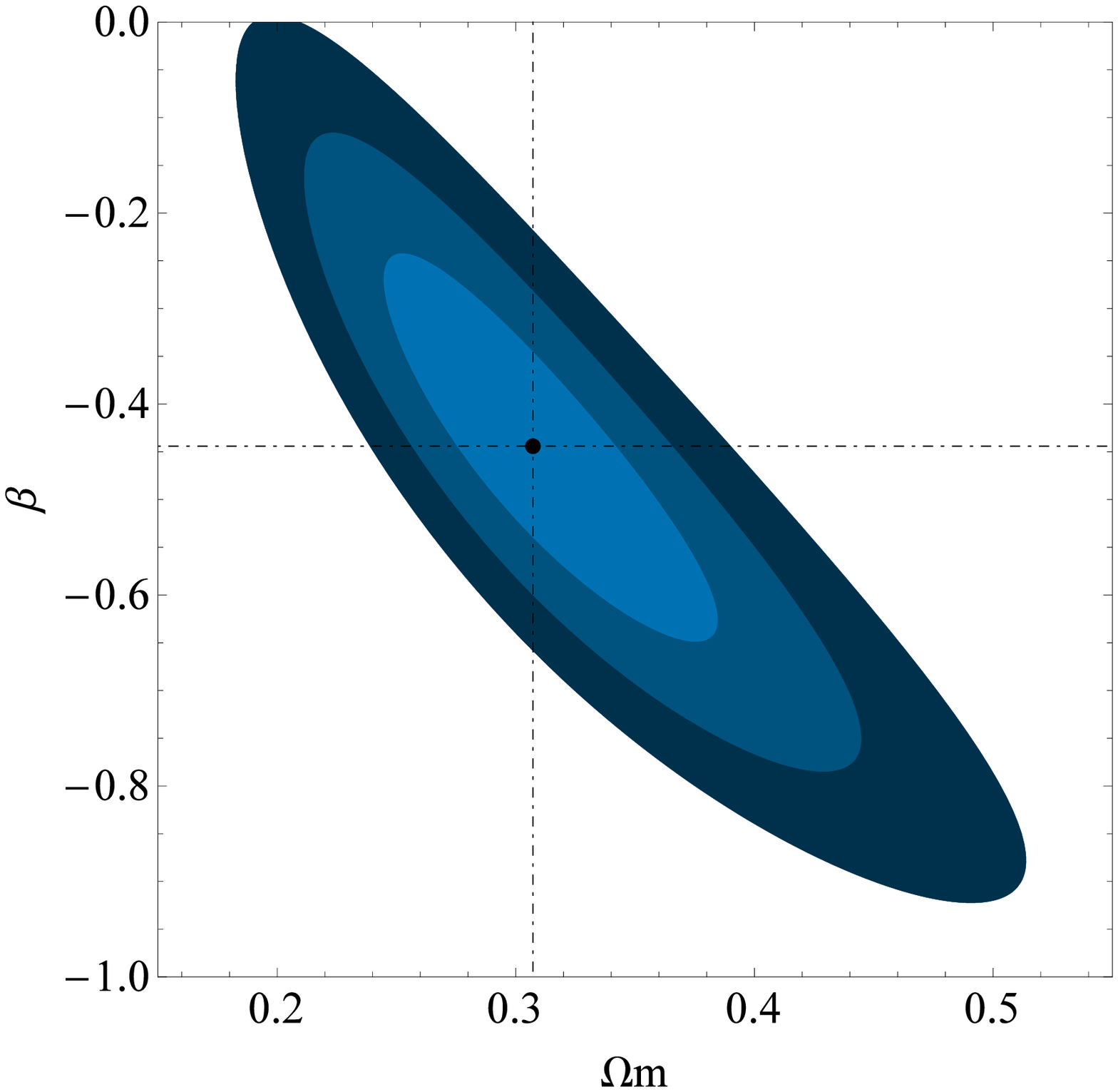}
        \caption{}
        \label{per:u-N}
    \end{subfigure}
        \caption{Plots of $1\sigma$ (light blue), $2\sigma$ (blue) and $3\sigma$
        (dark blue) confidence contours on $\alpha-\beta$ (a), $\alpha-\Omega_{m_{0}}$
        (b) and $\beta-\Omega_{m_{0}}$ (c) parameter spaces for SN Ia+BAO dataset
        in tachyonic chameleon dark energy scenario with $f(\phi)=f_0 \phi^\alpha$ and
        $V(\phi)=V_0 \phi^{\beta}$. The black dot in each plot shows the best fit point.}
    \label{f-v power}
\end{figure}

In this subsection we obtain observations bounds on the model
parameters where $f(\phi)$ and $V(\phi)$ are both in power-law forms
as follows:
\begin{align}
f(\phi) &= f_{0} \phi^\alpha,\nonumber\\
V(\phi) &= V_{0}\phi^\beta.
\end{align}
The results for the best fit values of the free parameters of the
model i.e. $\alpha$ and $\beta$  together with the present value of
$\Omega_{m_{0}}$ extracted from combined analysis  SN Ia + BAO are
presented in Table \ref{tab4}. In this case the contour plots of
various quantities for $68.3\%$, $99.4\%$ and $99.7\%$ confidence
level have been shown in Figure \ref{f-v power}. As in previous
cases we set $f_{0}=V_{0}=1$. One can clearly see that similar to
the previous cases this scenario is consistent with observations.
\begin{table}[H]
\begin{center}
\begin{tabular}{|c||c||c||c||c|}
\hline
Data&$\chi_{min}^2$&$\Omega_{m_{0}}$&$\alpha$&$\beta$\\
\hline
SN Ia +&584.08&0.30&1.23&-0.44\\
BAO&&&&\\
\hline
\end{tabular}
\caption{The value of $\chi_{min}^2$ and the best fit values of
model parameters $\Omega_{m_{0}}$, $\alpha$ and $\beta$ for the
third case.} \label{tab4}
\end{center}
\end{table}

\subsection{Exponential $f(\phi)$ and $V(\phi)$}
\begin{figure}[H]
    \centering
    \begin{subfigure}[H]{0.3\textwidth}
        \centering
        \includegraphics[width=50mm]{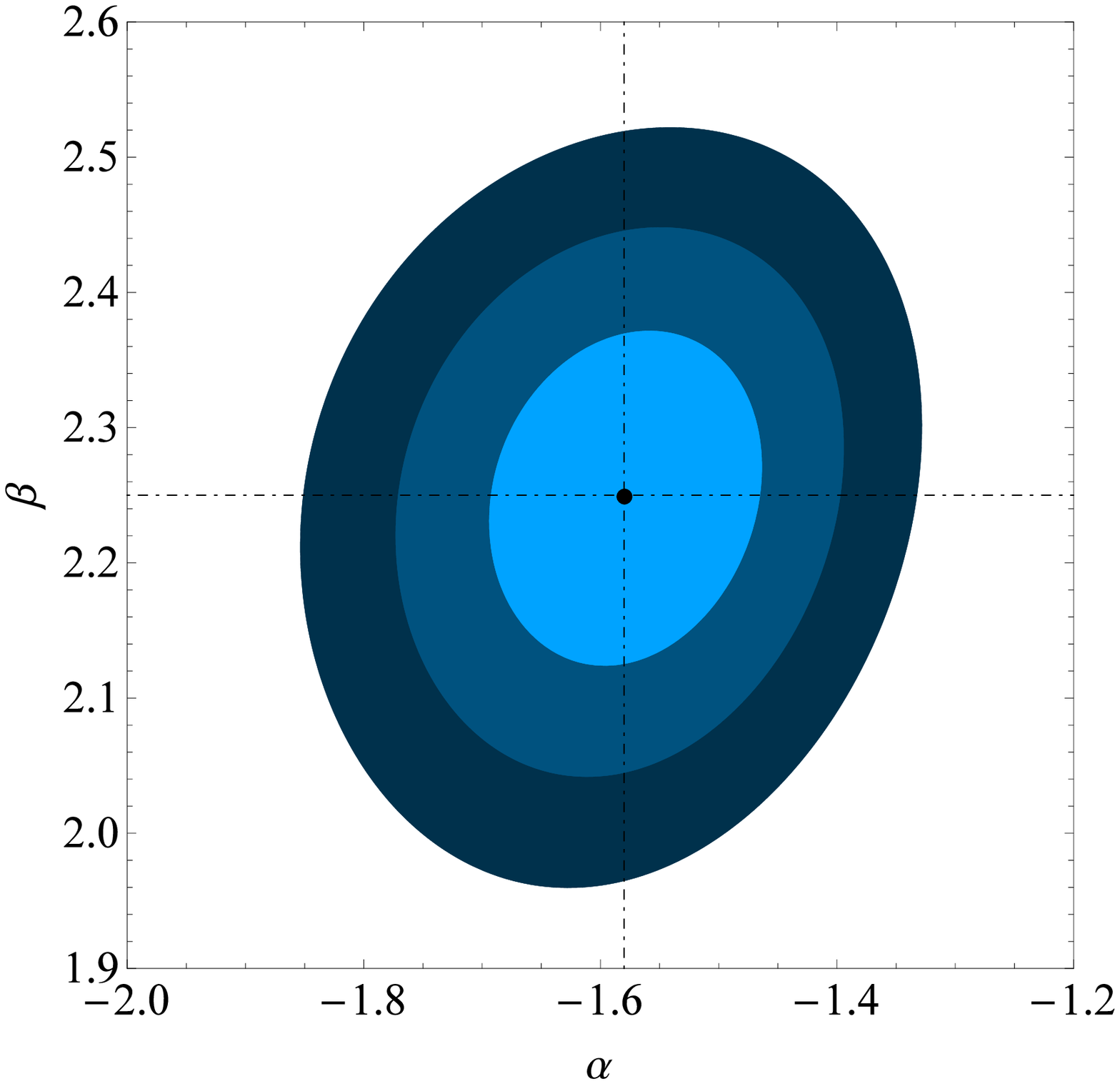}
        \caption{}
        \label{per:x-N}
    \end{subfigure}
    \begin{subfigure}[H]{0.3\textwidth}
        \centering
        \includegraphics[width=50mm]{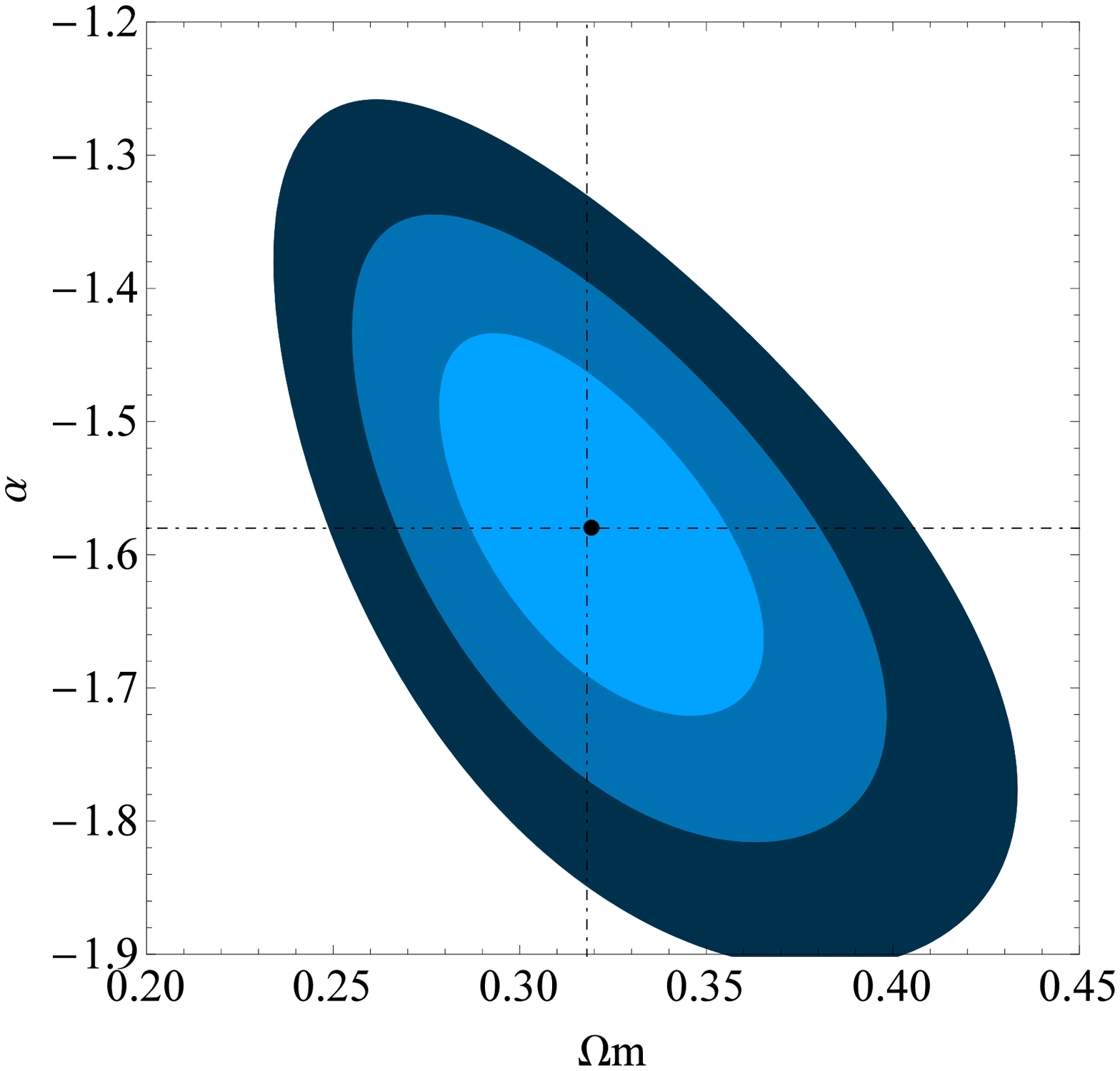}
        \caption{}
        \label{per:z-N}
    \end{subfigure}
       \begin{subfigure}[H]{0.3\textwidth}
        \centering
        \includegraphics[width=50mm]{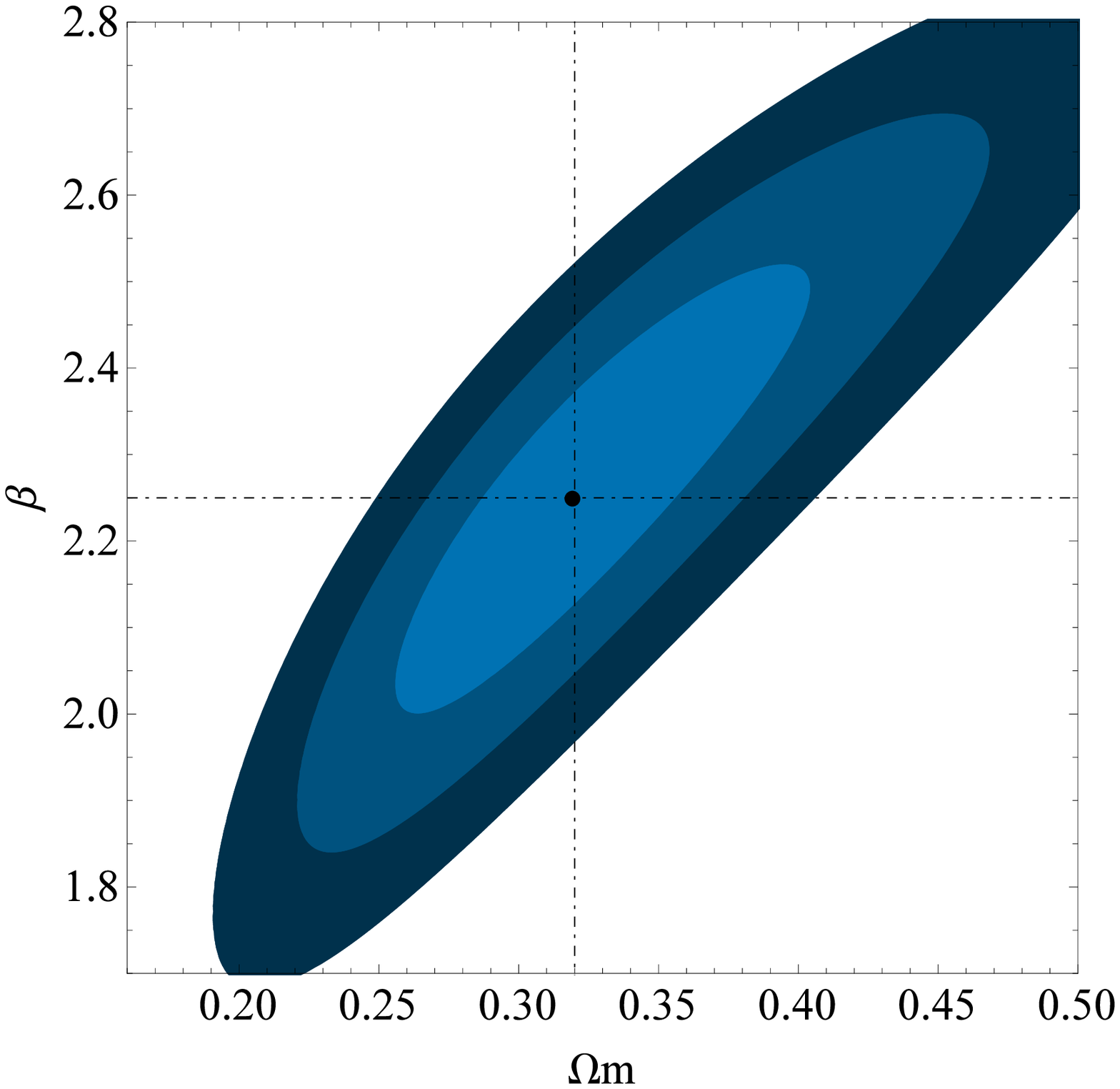}
        \caption{}
        \label{per}
    \end{subfigure}
        \caption{Plots of $1\sigma$ (light blue), $2\sigma$ (blue) and $3\sigma$ (dark blue)
        confidence contours on $\alpha-\beta$ (a), $\alpha-\Omega_{m}$ (b) and $\beta-\Omega_{m_{0}}$
        (c) parameter spaces for SN Ia+BAO dataset in tachyonic chameleon dark energy scenario with
        $f(\phi)=f_0 e^{\alpha\phi}$ and $V(\phi)=V_0 e^{\beta\phi}$. The black dot in each plot shows the best fit point.}
    \label{f-v exp}
\end{figure}

As a fourth case we obtain observational constraints from combined
datasets of SN Ia and BAO on the free parameters of the model where
the coupling function and the chameleon potential have exponential
forms given by:
\begin{align}
f(\phi) &= f_{0} e^{\alpha\phi},\nonumber\\
V(\phi) &= V_{0} e^{\beta\phi} .
\end{align}
The minimum value of $\chi^2$ as well as the best fit values of
$\Omega_{m_{0}}$, $\alpha$ and $\beta$ are presented in Table
\ref{tab5}. The contour plots of $\alpha$ versus $\beta$,
$\Omega_{m_{0}}$ versus $\alpha$ and $\Omega_{m_{0}}$ versus $\beta$
are shown in (a), (b) and (c) panels of Figure \ref{f-v exp}
respectively. For simplicity reasons we set $f_{0}=V_{0}=1$. As in
the previous cases these figures have been plotted for $1\sigma$,
$2\sigma$ and $3\sigma$ confidence levels. It is obvious from Figure
\ref{f-v exp} that the model is in agreement with observational data
from SN Ia in combination to BAO.\\
Before closing this subsection we mention that since the values of
the matter density parameter in all cases are very close to each
other, $\chi^{2}_{min}$'s are almost the same though the confidence
range at each case is different from the others and one can clearly
see such a difference in the contours.
\begin{table}[H]
\begin{center}
\begin{tabular}{|c||c||c||c||c|}
\hline
Data&$\chi_{min}^2$&$\Omega_{m_{0}}$&$\alpha$&$\beta$\\
\hline
SN Ia +&584.09&0.31&-1.58&2.25\\
BAO&&&&\\
\hline
\end{tabular}
\caption{The value of $\chi_{min}^2$ and the best fit values of
model parameters $\Omega_{m_{0}}$, $\alpha$ and $\beta$ for the
forth case.} \label{tab5}
\end{center}
\end{table}

\section{Conclusion}
In this paper we have used the latest observational data to
constrain the parameters of the tachyonic chameleon model of dark
energy. In our previous paper (Banijamali \& Solbi, 2017) we have
studied the dynamics of such a scenario and have found that this
model has the ability to alleviate the coincidence problem via the
mechanism of scaling attractors. Two important functions in our
analysis are tachyonic potential $V(\phi)$ and non-minimal coupling
function $f(\phi)$ in action (\ref{action}). In general we have
considered
two types of these functions i.e power-law and exponential forms.\\
We have fitted data from Type Ia supernova (SN Ia) and Baryon
Acoustic Oscillation (BAO) to constrain the present matter density
parameter $\Omega_{m_{0}}$, the parameter $\alpha$ in functional
form of $f(\phi)$ (the non-minimal coupling function $f$ is of the
form $f(\phi) \propto e^{\alpha\phi}$ or $f(\phi) \propto
\phi^\alpha$) and the parameter $\beta$ in $V(\phi)$ ( tachyonic
potential is assumed to be $V(\phi) \propto e^{\beta\phi}$ or
$V(\phi) \propto \phi^\beta$). For exponential $f(\phi)$ and
power-law $V(\phi)$ we have seen that positive $\alpha$ and negative
$\beta$ are favoured by the data while for power-law $f(\phi)$ and
exponential $V(\phi)$ both $\alpha$ and $\beta$ should be positive
in order to our model be compatible with observations. On the other
hand, when $f(\phi)$ and $V(\phi)$ are both power-law functions of
scalar field a positive value of $\alpha$ and a negative value of
$\beta$ are favoured while when these functions are both in
exponential forms then $\alpha$ is negative and $\beta$ is a
positive constant. We remark that in all four cases the value of
present matter density parameter
$\Omega_{m_{0}}$ is very close to the desired value.\\
In summay, the scenario of the tachyonic chameleon dark energy is
compatible with observations, for all examined scalar field
potential and non-minimal coupling functions.\\


\end{document}